%% file: main.tex
\documentclass[11pt]{article}

\usepackage{sbc-template}
\usepackage{anyfontsize}

\usepackage{color}

\usepackage[vlined,ruled,linesnumbered,portuguese]{algorithm2e}
\usepackage{amssymb}
\usepackage{amsmath}
\usepackage{natbib}
\usepackage{setspace}
\usepackage{float}
\usepackage{graphicx,url}
\usepackage{multirow}
\usepackage{algorithmic}
\usepackage[portuguese]{algorithm2e}
\usepackage[brazil]{babel}
\usepackage[utf8]{inputenc}
\SetKwFor{ParaCada}{para cada}{faça}{} 

\newcommand{\bif}[1]{\textbf{\emph{#1}}}

\makeatletter
	\def\hlinewd#1{%
		\noalign{\ifnum0=`}\fi\hrule \@height #1 \futurelet
	  \reserved@a\@xhline}
\makeatother

\title{Solving the minimum labeling global cut problem by mathematical programming} 

\author{Thiago Gouveia da Silva\inst{1,3,4}, Gilberto F. de Sousa Filho\inst{2,3}, 
Luiz Satoru Ochi\inst{3}, \\ Philippe Michelon \inst{4}, Serigne Gueye \inst{4}, Lucidio A. F. Cabral\inst{2} }

\address{Instituto Federal de Educação, Ciência e Tecnologia da Paraíba (IFPB)   \vspace*{-5pt}
\nextinstitute
  Centro de Informática -- Universidade Federal da Paraíba (UFPB)\\
  João Pessoa -- PB -- Brasil
 \nextinstitute
  Instituto de Computação -- Universidade Federal Fluminense (UFF)\\
  Niteroi -- RJ -- Brasil
 \nextinstitute
  Université d'Avignon et des Pays de Vaucluse (UAPV)\\
  Avignon -- França	
 \email{Corresponding author: thiago.gouveia@ifpb.edu.br}
}

\begin{document}

\maketitle

\newcommand{\XIPT}{\fontsize{11}{15}\selectfont }

\renewenvironment{abstract}{
  \vspace*{5pt}
  \begin{center}
  \bfseries ABSTRACT
	\end{center}
    \vspace*{-20pt}
    \hspace*{\parindent}
}

\newenvironment{keywords}{
    \fontsize{11}{15}\selectfont
		\noindent  \bfseries KEYWORDS:
}

\newenvironment{resumo1}{
    \begin{center}
		\vspace*{-5pt}
    \bfseries RESUMO
    \end{center}
    \vspace*{-20pt}
    \hspace*{\parindent}
}

\newenvironment{palchaves}{
    \fontsize{11}{15}\selectfont
    \noindent\bfseries PALAVRAS-CHAVE:
}

\vspace{3mm}


\begin{abstract}
\fontsize{11}{15}\selectfont{

Let $G=(V,E,L)$ be an edge-labeled graph such that $V$ is the set of vertices, $E$ is the set of edges, $L$ is the set of labels (colors) and each edge $e \in E$ has a label $l(e)$ associated; The goal of the minimum labeling global cut problem (MLGCP) is to find a subset $L' \subseteq L$ of labels such that $G' = (V, E', L \backslash L')$ is not connected and $|L'|$ is minimized. This work proposes three new mathematical formulations for the MLGCP as well as branch-and-cut algorithms to solve them. The computational experiments showed that the proposed methods are able to solve small to average sized instances in a reasonable amount of time.
}
\end{abstract}

\bigskip

\input{1_intro}
\input{2_related}

\input{3_form}
\input{5_result}

\input{6_conclu}

\bibliographystyle{sbpo}

\bibliography{refs}

\end{document}

%% file: 1_intro.tex
\newpage
\section{Introduction}
 
An increasing number of papers addressing problems defined on edge-labeled graphs (ELG) has been published in recent years. In this class of graphs, instead of weight, each edge has a label associated. For easy viewing, it is common to replace the labels with colors in these problems. For this reason the terms labels and colors will be used interchangeably in this work.

Probably, the most studied problem in this area is the minimum labeling spanning tree problem (MLSTP) \citep{Thiago:2014, Thiago:2015}, which consists of: given a ELG, find a spanning tree using the minimum number of colors. Other examples of problems defined on ELGs are the colorful traveling salesman problem (CTSP) \citep{Xiong:2007} and the maximum labeled matching problem (MLMP) \citep{Carrabs:2009}. Information on other problems defined on ELGs can be found in the review carried out by \citet{Granata:2013}.

\begin{figure}[h]
	\centering
		\includegraphics[scale=0.65]{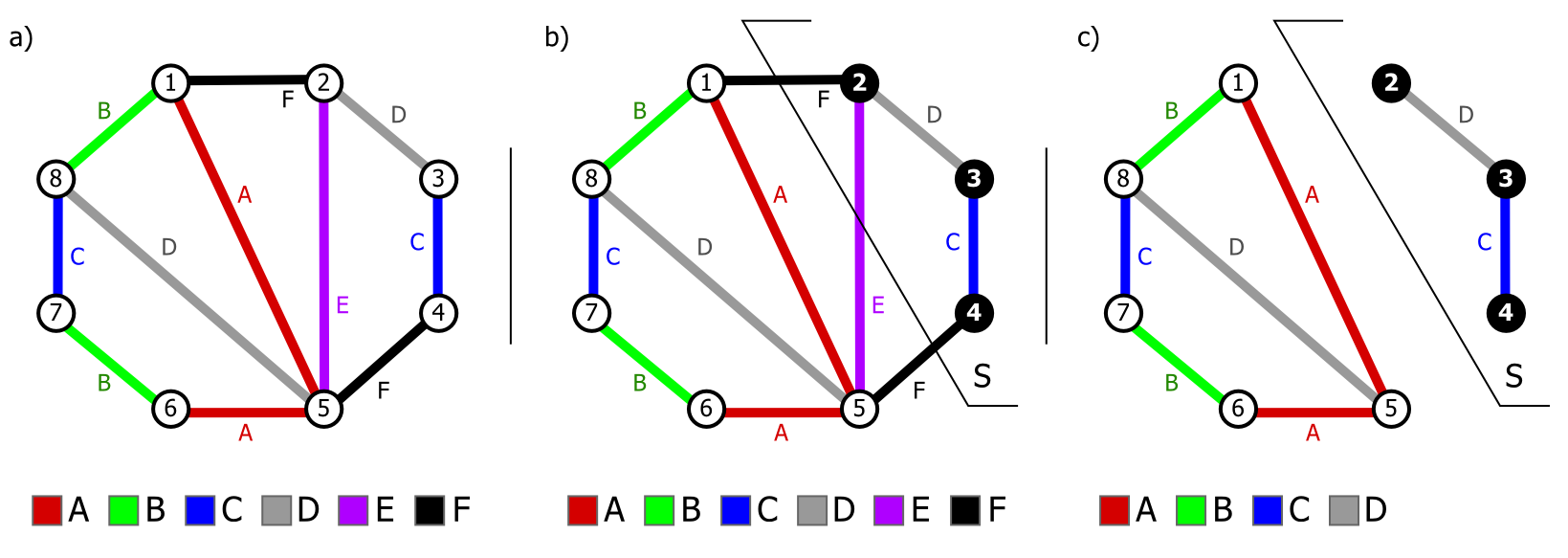}
	\caption{Example instance for the MLGCP.}
	\label{fig:exemplo}
\end{figure}

This work addresses the minimum labeling global cut problem (MLGCP). Given a ELG, the MLGCP aims to find the minimum number of labels such that the removal of all of its edges results in a disconnected graph. Figure \ref{fig:exemplo} illustrates the problem: Figure \ref{fig:exemplo}(a) presents the input graph, Figure \ref{fig:exemplo}(b) shows a cut composed by the labels $E$ and $F$, and Figure \ref{fig:exemplo}(c) shows that the graph without the edges of the cut is disconnected.

Formally, the MLGCP can be described as follows: given a non-oriented graph $G = (V, E, L)$, such that $V$ is the set of vertices, $E$ is the set of edges, $L$ the set of labels (colors) in $E$, and each edge $e \in E$ has a label $l(e)$ associated; the goal is to find a set of labels $L' \in L$ such that the graph $G' = (V, E', L \backslash L')$ is disconnected and $|L'|$ is minimized. That is, it uses the smaller number of labels  to disconnected the graph $G$.

In this work, we propose the first three mathematical formulations for the MLGCP, as well as exact techniques for solving them. Lastly, the computational experiments demonstrate that the proposed methods are able to solve small to medium-sized instances within reasonable computational time.

The subsequent sections are organized as follows: Section 2 presents a brief review of the literature on works related to MLGCP; Section 3 presents three mathematical formulations for the problem; Section 4 discusses the computational experiments and the results obtained; and Section 5 brings the final considerations and proposals for the continuity of the research.

%% file: 2_related.tex
\section{Related works}

 The minimum labeled s-t cut problem (MLstCP) is problem strongly related to  the MLGCP. Given an ELG $G$ and two vertices $s, t \in V$, the MLstCP aims to remove the minimum number of labels from $G$ such that $s$ and $t$ are disconnected in the resulting graph. The MLstCP was proposed by \citet{Jha:2002}, which proved the problem is NP-hard.

The MLstCP was motivated by the generation and analysis of attack graphs \citep{Jha:2002, shey:2002, shey:2003}, a system security problem in which the vertices represent the possible states of the attacker (intruder), $s$ represents the initial state, $t$ the success of the attack; and a edge $(u,v)$ indicates that the attacker has changed from state $u$ to $v$ using the attack $k$. Since it cripples each type of attack $k$ involves a cost, the goal is to prevent the attacker from reaching the state $t$ by disabling as lesser types of attack as possible.


\citet{Fellows:2009} studied the MLstCP with respect to their parameterized complexity, demonstrating that the problem is W[2]-Hard in graphs with width on the way (\textit{pathwidth}) less or equal to $3$, and W[1]-Hard for graphs with \textit{pathwidth} less or equal to $4$ even when the problem is bounded by the number of edges of the solution.

Several approximate algorithms were proposed for the MLstCP \citep{Zhang:2011,tang:2012,zhang:2014}. The best results in this area were obtained by \citet{zhang:2014}, demonstrating that the problem is $l_{\text{max}}-$ approximate and $f_{\text{max}}-$ approximate, being $l_{\text{max}}$ the length of the longest path $s-t$ and $f_{\text{max}}$ the largest number of edges a single label has.

The approximate algorithms proposed by \citet{tang:2012} were based on two mathematical formulations for the MLstCP. To the best of our knowledge, these are the only proposed formulations for the problem until the present date. Let $z_l$ be a variable that takes value $1$ if and only if the label $l \in L$ participate in the solution,  $\mathcal{P}_{st}$ be the set of all simple paths between $s$ and $t$, and be $L(P)$, $P \in \mathcal{P}_{st}$,  the set of labels used by the edges of $P$, the formulation based on labeled paths follows:

\begin{align}
\mathrm{Minimize}
    \sum_{l \in L} z_l \label{lp2:fo}
\end{align}

\begin{align}
	\text{s.t.} \quad \sum_{l \in L(P)} z_{l} \geq 1, && \forall P \in \mathcal{P}_{st}, \label{lp2:eq1} \\
	z_{l} \in \{ 0,1 \}, && \forall l\in L. \label{lp2:eq2}
\end{align}

The works from \citet{coudert:2007} and \citet{coudert:2014} address the MLGCP and the MLstCP with the goal of measuring the ability of a network to remain connected when link pools are at shared risk. For example, on a Wi-Fi network, an attacker can knock down all links at a certain frequency by adding a strong noise signal to this. Another case occurs when two links use the same duct in part of the way.

Lastly, \citet{zhang:2014} shows that the MLGCP can be solved in polynomial time if the input graph is planar, has a limite tree width  or a small value of $f_{\text{max}}$. Heuristics were proposed for the MLGCP by \citep{SILVA201823} and \citep{Bordini:2017}.

%% file: 3_form.tex
\section{Mixed integer linear programming models}
\label{sec:form}

In this section, we present three models of mixed integer linear programming (MILP) to solve the MLGCP. Besides, we reformulate the first model by reducing the number of decision variables and propose \textit{branch-and-bound} (B\&B) and \textit{branch-and-cut} (B\&C) strategies to solve the proposed formulations.

\subsection{Partition based model}

The first model proposed, denominated PART, aims to partition the set of vertices into two groups: $S$ and $\overline{S} = V \backslash S$. For any set $S \subset V$, $S \neq \emptyset$, the removal of all edges with one end in $S$ and another in $\overline{S}$ disconnects the graph.  Let $e_{ij}$ be an edge with ends in vertices $i$ and $j$; $z_l$, $\forall l \in L$, be a variable that assumes value $1$ if and only if the label $l$ participate of the cut; $x_e$, $\forall e \in E$, be a variable that assumes value $1$ if and only if the edge $e$ participate in the cut; and $w_v$, $\forall v \in V$ be a variable that assumes value $1$ if and only if the vertex $v$ is part of the set $S$; the PART model is described in expressions (\ref{ccut:fo}-\ref{ccut:eq7}):

\begin{align}
\text{PART} = \mathrm{Minimize}
    \sum_{l \in L} z_l \label{ccut:fo}
\end{align}

\begin{align}
	\text{s.t.} \quad 1 \leq \sum_{v \in V} w_v < |V|, \label{ccut:eq1} \\
	z_{l(e)} \geq x_e, && \forall e \in E, \label{ccut:eq2} \\[5pt]
	x_{e_{ij}} \geq w_i - w_j, && \forall e_{ij} \in E, \label{ccut:eq3} \\[5pt]
	x_{e_{ij}} \geq w_j - w_i, && \forall e_{ij} \in E, \label{ccut:eq4} \\[5pt]
	z_{l} \geq 0, && \forall l\in L, \label{ccut:eq5} \\[5pt]
	x_{e} \geq 0, && \forall e\in E, \label{ccut:eq6} \\[5pt]
  w_{v} \in \{ 0,1 \}, && \forall v\in V. \label{ccut:eq7} 
\end{align}

The objective function (\ref{ccut:fo}) minimizes the number of labels necessary to disconnect the graph; the constraint (\ref{ccut:eq1}) forces that $S \subset V$ and $S \neq \emptyset$; the set of constraints (\ref{ccut:eq2}) binds the label variables to the edge variables; the inequalities (\ref{ccut:eq3}) e (\ref{ccut:eq4}) activate the cutting edges; and the restrictions (\ref{ccut:eq5}-\ref{ccut:eq7})) define the domain of the variables.

Let L$_c$ be the cutting set of a \text{PART} solution. Figure \ref{fig:exemplo}(c) presents a possible solution for \text{PART}, where $S=\{2, 3, 4\}$, $\overline{S}=\{1, 5, 6, 7, 8\}$, $z_E=1$ and $z_F=1$, so L$_c = \{E, F\}$ and $|$L$_c| = 2$, which is the optimal solution value of this instance.   

Note that, by the constraints (\ref{ccut:eq3}-\ref{ccut:eq4}), we have $z_e \geq x_e \geq |w_i - w_j|$, which is equivalent to $z_e \geq |w_i - w_j|$. In such case, the edge variables could be eliminated from the formulation. Besides, we can eliminate the symmetry of the solutions by choosing an arbitrary vertex to participate in $S$. The expressions (\ref{ccut2:fo}-\ref{ccut2:eq6}) presents the reformulated model, called PART$_2$:

\begin{align}
\text{PART}_2 = \mathrm{Minimize}
    \sum_{l \in L} z_l \label{ccut2:fo}
\end{align}

\begin{align}
	\text{s.t.} \quad \sum_{v \in V} w_v < |V|, \label{ccut2:eq1} \\
	w_1 = 1, \label{ccut2:eq2} \\[5pt]
 	z_{l(e_{ij})} \geq w_i - w_j, && \forall e_{ij} \in E, \label{ccut2:eq3} \\[5pt]
	z_{l(e_{ij})} \geq w_j - w_i, && \forall e_{ij} \in E, \label{ccut2:eq4} \\[5pt]
	z_{l} \geq 0, && \forall l\in L, \label{ccut2:eq5} \\[5pt]
  w_{v} \in \{ 0,1 \}, && \forall v\in V. \label{ccut2:eq6} 
\end{align}

\subsection{Vertex clustering based model}


The next model we propose is based on the problem of clustering vertices by edge editions. Given a complete graph $K_n$ with costs (positive or negative) associated with the edges, the goal is to remove edges (paying the cost of removal) so that the resulting graph is a partitioning of cliques. In other words, we obtain a graph in which each group of vertices forms a clique and the edges between two cliques form a disconnecting set. To transform the input graph $G$ on a $K_n$, for each pair of vertices $i,j$ such that $e_{ij} \notin E$, we added an edge to $G$ with removal cost zero.

Let $\overline{E}$ be the set of all edges that do not exist in $G$, so that $E \cup \overline{E}$ create a complete graph; and let $x_e$, $\forall e \in E \cup \overline{E}$ and $z_l$, $\forall l \in L$ decision variables as defined for the PART formulation. The expressions (\ref{ccut3:fo}-\ref{ccut3:eq7}) present the P3E model. The objective function (\ref{ccut3:fo}) minimizes the cost of the solution labels; the set of constraints (\ref{ccut3:eq1}) binds the edge variables to the color variables; the set of equations (\ref{ccut3:eq2}-\ref{ccut3:eq4}) is the classic set of P$_3$ elimination (paths with three vertices); the inequality (\ref{ccut3:eq5}) ensures the solution is not empty; and the expressions (\ref{ccut3:eq6}-\ref{ccut3:eq7}) define the domain of decision variables.

\begin{figure}[h]
	\centering
		\includegraphics[scale=0.45]{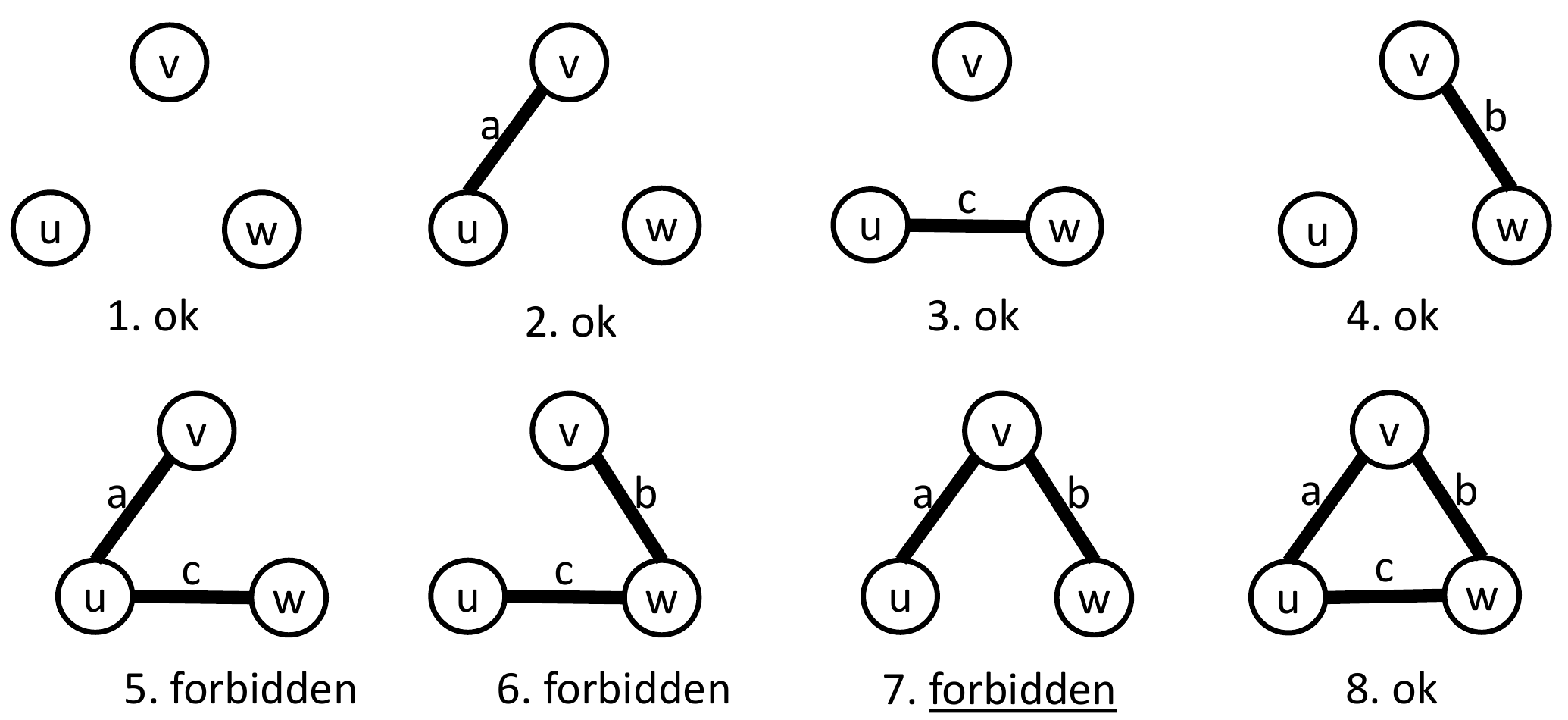}
	\caption{Edge configurations prohibited and allowed by inequalities (\ref{ccut3:eq2} - \ref{ccut3:eq4}) of \text{P3E}.}
	\label{fig:p3}
\end{figure}

According to \citet{Grotschel1990}, if $A$ is partitioning in cliques $a=uv$ and $b=vw$ are two edges in $A$ with the vertex $v$ in common, then the edge $c=uw$ must also be in $A$. The Figure \ref{fig:p3} illustrates allowed and forbidden edge settings for partitioning in cliques. 

Therefore, since the input graph is connected, the correctness of the model is based on the fact that $G$ is a partitioning of cliques if and only if P$_3$ is not a sub-graph of $G$, being P$_3$ a path formed by three vertices. Finally, given the size of the set of P$_3$ elimination constraints, we propose the resolution of the model by B\&C. The proposed algorithm starts the model without any elimination restriction of P$_3$ and, for any solution found, check by enumeration if any restriction of this set has been violated. 

\begin{align}
\text{P3E} = \mathrm{Minimize}
    \sum_{l \in L} z_l \label{ccut3:fo}
\end{align}

\begin{align}
	\text{s.t.} \quad z_{l(e)} \geq x_e, && \forall e \in E, \label{ccut3:eq1} \\[5pt]
	+ x_{ij} + x_{jk} - x_{ki} \geq 0, && \forall i,j,k \in V, i \neq j, j \neq k, k \neq i, \label{ccut3:eq2} \\[5pt]
	+ x_{ij} - x_{jk} + x_{ki} \geq 0, && \forall i,j,k \in V, i \neq j, j \neq k, k \neq i, \label{ccut3:eq3} \\[5pt]
	- x_{ij} + x_{jk} + x_{ki} \geq 0, && \forall i,j,k \in V, i \neq j, j \neq k, k \neq i, \label{ccut3:eq4} \\[5pt]		
  \sum_{e \in E} x_e \geq 1, \label{ccut3:eq5} \\	
	z_{l} \geq 0, && \forall l\in L, \label{ccut3:eq6} \\[5pt]
	x_{e} \in \{ 0,1 \}, && \forall e\in E \cup \overline{E}. \label{ccut3:eq7}
\end{align}

\subsection{Model based on elimination of trees}

Given that a spanning tree is a minimal connected graph with respect to its number of edges, the elimination of all spanning trees of $G$ results in a disconnected graph. Let $\mathcal{T}$ be the set of all spanning trees of the graph $G$; $L(T)$, $T \in \mathcal{T}$, be the set of labels used by the edges of $T$; and $z_l$, $\forall l \in L$, be decision variables as defined for the PART formulation, the expressions (\ref{t:fo}-\ref{t:eq2}) present the EAC model:

\begin{align}
\text{EAC} = \mathrm{Minimize}
    \sum_{l \in L} z_l \label{t:fo}
\end{align}

\begin{align}
	\text{s.t.} \quad \sum_{l \in L(T)} z_{l} \geq 1, && \forall T \in \mathcal{T}, \label{t:eq1} \\
	z_{l} \in \{ 0,1 \}, && \forall l\in L. \label{t:eq2}
\end{align}

The objective function (\ref{t:fo}) minimizes the cost of the solution labels; the exponential set of constraints (\ref{t:eq1}) ensures the disconnection of the solution graph by forbidding all spanning trees of $G$; and the constraints (\ref{t:eq2}) define the domain of the variables $z$. 

\begin{figure}[h]
	\centering
		\includegraphics[scale=0.65]{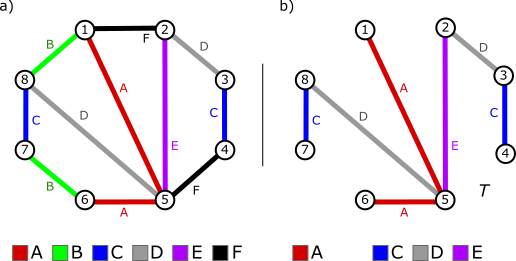}
	\caption{Spanning tree contained in a example instance of the MLGCP.}
	\label{fig:tree}
\end{figure}

Figure \ref{fig:tree}(a) shows a spanning tree $T$ obtained from the input instance presented in Figure \ref{fig:tree}(b). Then, $T$ can not be allowed to exist to ensure that the graph $G$ be disconnected. In this sense, the addition of the inequality

\[
z_A + z_C + z_D + z_E \geq 1
\]

\noindent ensures that at least one of the labels of $T$ is part of the solution cut.

Once the number of generating trees of a graph is $O(n^{n-2})$, it is not practical to add the whole set of constraints (\ref{t:eq1}). For this, the B\&C algorithm proposed starts the model without any constraints and, for any integer solution found, verifies whether it is possible to construct a spanning tree with the labels that remain in the graph. In positive case, it adds the constraint that removes this tree from the model. We use the heuristic MVCA \citep{Krumke:1998} to find spanning trees using as few labels as possible.

Additionally, we propose a separation heuristic for the tree elimination constraints. Let $\overline{z}_l$ be the value of the linear relaxation of the variable $z_l$ and $H$ be an empty graph with all the vertices of $G$: select the label $l$ with lower value $\overline{z}_l$ and add all edges with label $l$ to $H$; repeat this procedure until $H$ is connected; if the sum of the values $\overline{z}_l$ of the labels added to $H$ is less than $1$, this cut is violated.

%% file: 5_result.tex
\section{Computational results}

All the methods studied in this work were developed in C++, using the compiler g++ version 4.6.3 (compile option -O3), and the MIP-solver CPLEX version 12.04. The computational experiments were carried out on a computer with a \textit{Intel Core} I7 (64 bits) processor, CPU with 4 cores of 2,93 GHz, in addition to 8 GB of RAM and running the operating system Linux Ubuntu 14.04. The experiments were carried out with $1$ \textit{thread} and time limit of one hour.

For the experiments performed, we have used the ELGs generated at random by \cite{Cerulli:2005}. For each $n=|V|= 50, 100, 200$, the instances have 120 problem-graphs. Each set consists in $12$ groups of instances with different dimensions, each group being defined by a combination between $|L|=\{\frac{n}{4};\frac{n}{2};n;\frac{5n}{4}\}$ and $d=\{ld=0,2; md=0,5; hd=0,8\}$, \textit{d} representing a density of the graph. Each group has $10$ different instances with the same dimension.

Tables \ref{tab:u50} and \ref{tab:u200} present the comparison of the computational results of the methods proposed in this work, each row representing a group with $10$ instances of same dimension. In these tables we have the first group of columns representing the groups of tested instances, where the column \textit{UB} stands for the mean of the best integer solutions obtained by the methods, while the remaining columns are divided into three groups: \textit{PART$_2$}, \textit{P3E} and \textit{EAC}, represented the compared methods. Each method has the column \textit{O}, indicating the number of instances for which the method reached an optimal solution; \textit{t(s)} indicating the average execution time in seconds; \textit{gap} representing the mean percentage of the difference between the LB and UB obtained by the method; and \textit{gapr} indicating the average percentage of the difference between the linear relaxation of the model and its LB. 

We can observe from Table \ref{tab:u50} that the models \textit{PART$_2$} and \textit{EAC} reached the optimum in all instances. Although the model \textit{P3E} failed for only one instance, this model presented very high computational times as in the instances $\{|V|=50; hd; |L|=50\}$ with average 1280 seconds to be solved, while \textit{PART$_2$} solved these instances in only 1.48 seconds on average. The method \textit{EAC} obtained the best values for the linear relaxation in all instances, still obtaining the best computational time for instances with small $d$ and $|L|$.

The method \textit{P3E} was not inserted in Table \ref{tab:u100} because it could not reach any integer solutions within one hour of execution. The strategy \textit{EAC} continues to find the best values of linear relaxation in all instances. However, only for instances with $|L|=25$ that its computational time is smaller than the \textit{PART$_2$} method. All instances were resolved by the \textit{PART$_2$} method. The \textit{EAC} method failed to solve many instances, such as the group $\{|V|=100; md; |L|=125\}$. It is possible to observe a degradation in the \textit{EAC} method with the growth in the values of $|L|$ and $d$.

Lastly, Table \ref{tab:u200} shows, for $|V|=200$, that the \textit{EAC} method continues to find the best results for the linear relaxation. However, it failed to in three groups of instances marked with "*", because they did not obtain integer solutions within one hour of execution. The \textit{PART$_2$} method, even getting worse linear relaxations in all instances, not solved only 17 instances out of 120 with $|V|=200$, failing in instances with $|L| \geq 200$ and $d =\{md, hd\}$.   

\input{tabelas/unicost.50} 

\input{tabelas/unicost.100} 

\input{tabelas/unicost.200} 

%% file: tabelas/unicost.50.tex
\begin{table}[htb]
	\centering
	\caption{Computational results for instances with $|V| = 50$.}
  \vspace{3mm}	
  \setlength{\tabcolsep}{5.6pt}
	
	\footnotesize
	\begin{tabular}{r l r c r r r r c r r r r c r r r r }

	\hline
	\multicolumn{3}{c}{\bif{Instances}} & & \multicolumn{4}{c}{\bif{P3E}} & & \multicolumn{4}{ c }{\bif{PART$_2$}} & & \multicolumn{4}{ c }{\bif{EAC}}\\
	
	\cline{1-3} \cline{5-8} \cline{10-13} \cline{15-18}
	
	\bif{$|L|$} & \bif{d} & \bif{UB} & & \bif{O} &
	\multicolumn{1}{c}{\bif{t (s)}}  &
	\multicolumn{1}{c}{\bif{gap}}  &
	\multicolumn{1}{c}{\bif{gapr}}  & & \bif{O} &
	\multicolumn{1}{c}{\bif{t (s)}}  &
	\multicolumn{1}{c}{\bif{gap}}  &
	\multicolumn{1}{c}{\bif{gapr}}  & & \bif{O} &
	\multicolumn{1}{c}{\bif{t (s)}}  &
	\multicolumn{1}{c}{\bif{gap}}  &
	\multicolumn{1}{c}{\bif{gapr}}  \\

	\hlinewd{1.5pt}	
	
		& ld & 2,5 & & 10 & 843,4 & 0 & 67,4 & & 10 & 0,05 & 0 & 56,0 & & 10 & 0,006 & 0 & 6 \\
12  & md & 7,4 & & 10 & 947,0 & 0 & 90,4 & & 10 & 0,08 & 0 & 77,8 & & 10 & 0,001 & 0 & 22,0 \\
		& hd & 9,8 & & 10 & 596,7 & 0 & 69,9 & & 10 & 0,05 & 0 & 35,3 & & 10 & 0,001 & 0 & 24,2 \\

  &    \\

		& ld & 2,7 & & 10 & 1015,9 & 0 & 74,6 & & 10 & 0,183 & 0 & 38,1 & & 10 & 0,009 & 0 & 8,6 \\
25	& md & 9,9 & & 10 & 997,5 & 0 & 93,4 & & 10 & 0,31 & 0 & 81,7 & & 10 & 0,04 & 0 & 38,9 \\
		& hd & 15,5 & & 10 & 1511,2 & 0 & 96,1 & & 10 & 0,31 & 0 & 87,9 & & 10 & 0,05 & 0 & 31,7 \\

  &    \\

		& ld & 2,8 & & 9 & 1399,9 & 8,9 & 73,6 & & 10 & 0,17 & 0 & 37,5 & & 10 & 0,04 & 0 & 13,7 \\
50	& md & 11,6 & & 10 & 805,2 & 0 & 94,8 & & 10 & 0,82 & 0 & 85,1 & & 10 & 0,85 & 0 & 40,3 \\
		& hd & 21,3 & & 10 & 1280,9 & 0 & 97,2 & & 10 & 1,48 & 0 & 90,8 & & 10 & 3,1 & 0 & 45,9 \\

  &    \\

		& ld & 2,8 & & 10 & 1124,9 & 0 & 73,8 & & 10 & 0,13 & 0 & 38,0 & & 10 & 0,05 & 0 & 11,1 \\
62	& md & 12,1 & & 10 & 715,0 & 0 & 95,1 & & 10 & 1,1 & 0 & 85,8 & & 10 & 4,4 & 0 & 45,5 \\
		& hd & 22,7 & & 10 & 809,4 & 0 & 97,7 & & 10 & 1,8 & 0 & 91,0 & & 10 & 12,7 & 0 & 58,6 \\

	\hline	
	\end{tabular}
	\label{tab:u50}
\end{table}

%% file: tabelas/unicost.100.tex
\begin{table}[htb]
	\centering
	\caption{Computational results for instances with  $|V| = 100$.}
  \vspace{3mm}	
  \setlength{\tabcolsep}{5,6pt}
	
	\footnotesize
	\begin{tabular}{r l r c r r r r c r r r r }

	\hline
	\multicolumn{3}{c}{\bif{Instances}} & & \multicolumn{4}{ c }{\bif{PART$_2$}} & & \multicolumn{4}{ c }{\bif{EAC}}\\
	
	\cline{1-3} \cline{5-8} \cline{10-13}
	
	\bif{$|L|$} & \bif{d} & \bif{UB} & & \bif{O} &
	\multicolumn{1}{c}{\bif{t (s)}}  &
	\multicolumn{1}{c}{\bif{gap}}  &
	\multicolumn{1}{c}{\bif{gapr}}  & & \bif{O} &
	\multicolumn{1}{c}{\bif{t (s)}}  &
	\multicolumn{1}{c}{\bif{gap}}  &
	\multicolumn{1}{c}{\bif{gapr}}  \\

	\hlinewd{1,5pt}

		&	ld & 6,2 & & 10 & 0,9 & 0 & 83,2 & & 10 & 0,06 & 0 & 37,9 \\
25	&	md & 16,5 & & 10 & 0,7 & 0 & 91,1 & & 10 & 0,09 & 0 & 35,6 \\
		&	hd & 21 & & 10 & 0,5 & 0 & 86,3 & & 10 & 0,1 & 0 & 39,2 \\

  &    \\

		&	ld & 6,8 & & 10 & 2,5 & 0 & 83,7 & & 10 & 0,4 & 0 & 37,7 \\
50	&	md & 22,2 & & 10 & 5,7 & 0 & 93,5 & & 10 & 6,5 & 0 & 55,1 \\
		&	hd & 33,1 & & 10 & 4,8 & 0 & 95,8 & & 10 & 6,5 & 0 & 48,3 \\

  &    \\

		&	ld & 7,2 & & 10 & 2,1 & 0 & 84,3 & & 10 & 10,9 & 0 & 41,5 \\
100	&	md & 26,5 & & 10 & 9,3& 0 & 95,3 & & 5 & 2012,7 & 8,4 & 63,9 \\
		&	hd & 45,2 & & 10 & 22,4 & 0 & 96,4 & & 2 & 3071,5 & 10,3 & 61,8 \\

  &    \\

		&	ld & 7,2 & & 10 & 3,1 & 0 & 84,1 & & 10 & 75,8 & 0 & 38,0 \\
125	&	md & 27,1 & & 10 & 36,1 & 0 & 95,0 & & 1 & 3409,0 & 17,8 & 55,7 \\
		&	hd & 45,2 & & 10 & 22,4 & 0 & 96,4 & & 2 & 3071,8 & 10,3 & 61,8 \\

	\hline	
	\end{tabular}
	\label{tab:u100}
\end{table}

%% file: tabelas/unicost.200.tex
\begin{table}[htb]
	\centering
	\caption{Computational results for instances with  $|V| = 200$.}
  \vspace{3mm}	
  \setlength{\tabcolsep}{5.6pt}
	
	\footnotesize
	\begin{tabular}{r l r c r r r r c r r r r }

	\hline
	\multicolumn{3}{c}{\bif{Instances}} & & \multicolumn{4}{ c }{\bif{PART$_2$}} & & \multicolumn{4}{ c }{\bif{EAC}}\\
	
	\cline{1-3} \cline{5-8} \cline{10-13}
	
	\bif{$|L|$} & \bif{d} & \bif{UB} & & \bif{O} &
	\multicolumn{1}{c}{\bif{t (s)}}  &
	\multicolumn{1}{c}{\bif{gap}}  &
	\multicolumn{1}{c}{\bif{gapr}}  & & \bif{O} &
	\multicolumn{1}{c}{\bif{t (s)}}  &
	\multicolumn{1}{c}{\bif{gap}}  &
	\multicolumn{1}{c}{\bif{gapr}}  \\

	\hlinewd{1.5pt}

		& ld & 13,2 & & 10 & 34,4 & 0 & 93,7 & & 10 & 5,7 & 0 & 57,4 \\
50	& md & 32,7 & & 10 & 9,7 & 0 & 97,1 & & 10 & 9,4 & 0 & 48,5 \\
		& hd & 43,3 & & 10 & 6,9 & 0 & 97,5 & & 10 & 2,5 & 0 & 43,4 \\

&    \\

		& ld & 15 & & 10 & 188,3 & 0 & 94,7 & & 9 & 1219,8 & 3,1 & 55,0 \\
100	& md & 45,4 & & 10 & 699,5 & 0 & 98,3 & & 0 & 3600 & 15,5 & 62,9 \\
		& hd & 68,8 & & 10 & 238,9 & 0 & 98,7 & & 0 & 3600 & 9,4 & 62,8 \\

&    \\

		& ld & 15,9 & & 10 & 614,4 & 0 & 94,0 & & 0 & 3600 & 45,4 & 61,6 \\
200	& md & 54,1 & & 6 & 2066,0 & 11,1 & 98,2 & & 0 & 3600 & 32,3 & 56,9 \\
		& hd & 93,8 & & 7 & 2051,9 & 14,3 & 98,9 & & 0 & 3600 & * & * \\

&    \\

		& ld & 16,1 & & 10 & 691,6 & 0 & 93,9 & & 0 & 3600 & 42,7 & 56,4 \\
250	& md & 56,5 & & 3 & 2990,2 & 35,6 & 98,1 & & 0 & 3600 & * & * \\
		& hd & 93,8 & & 7 & 2051,3 & 14,3 & 98,9 & & 0 & 3600 & * & * \\

	\hline	
	\end{tabular}
	\label{tab:u200}
\end{table}

%% file: 6_conclu.tex
\section{Conclusions and Future Works}

In this work, to the best of our knowledge, we propose the first three mathematical formulations for the  minimum labeling global cut problem. The first formulation, called \textit{PART$_2$}, proposes the generation of two partitions of the set $V$ and the accounting of the labels belonging to the cut between these two sets. In the model \textit{P3E}, the partitioning of the set $V$ is achieved by the clustering of the vertices through the classical $P_3$ prohibition inequalities. And the third model, called \textit{EAC}, presents a set of exponential constraints to prohibit the existence of any labeled spanning tree of the set $V$, ensuring a disconnected set. Since the \textit{P3E} and \textit{EAC} models have a large number of inequalities, we propose separation heuristics for these cuts, implementing strategies of \textit{branch-and-cut} for these models. 

Lastly, we performed computational tests which demonstrated that the proposed methods are able to solve small to medium-sized instances with reasonable computational effort. The \textit{P3E} method presented a poor performance due to the large amount of constraints added to the model. The \textit{PART$_2$} model has obtained the best performance among the three methods proposed, even though it had a very poor linear relaxation in comparison to the \textit{EAC} method. This performance is due to the linear number of constraints of the model and to the \textit{branch-and-bound} on the variable $w_i$, which has the linear dimension $|V|$, and is a very strong decision for the \textit{branch}, indicating in which partition the vertex $i$ should be in solution.

As future works we intend to create new separation methods to improve the linear relaxation of the \textit{PART$_2$} model and increase the convergence of the strategy \textit{EAC}, adding cuts of the polyhedral hull of these models.